\documentclass{article}
\usepackage{spconf,amsmath,graphicx}
\usepackage{xcolor}
\usepackage{booktabs}
\usepackage{hyperref}
\usepackage{caption}
\usepackage{verbatim,enumitem}

\title{Universal Paralinguistic Speech Representations using Self-Supervised Conformers}
\name{Joel Shor$^1$, Aren Jansen$^2$, Wei Han$^2$, Daniel Park$^2$, Yu Zhang$^2$}
\address{Verily Life Sciences, Boston, USA$^1$ and Mountain View, California, USA$^2$ \\
joelshor@verily.com}

\begin{document}

\ninept
\sloppy
\maketitle
\begin{abstract}
Many speech applications require understanding aspects beyond the words being spoken, such as recognizing emotion, detecting whether the speaker is wearing a mask, or distinguishing real from synthetic speech. In this work, we introduce a new state-of-the-art paralinguistic representation derived from large-scale, fully self-supervised training of a 600M+ parameter Conformer-based architecture.  We benchmark on a diverse set of speech tasks and demonstrate that simple linear classifiers trained on top of our time-averaged representation outperform nearly all previous results, in some cases by large margins.  Our analyses of context-window size demonstrate that, surprisingly, 2 second context-windows achieve 96\% the performance of the Conformers that use the full long-term context on 7 out of 9 tasks. Furthermore, while the best per-task representations are extracted internally in the network, stable performance across several layers allows a single universal representation to reach near optimal performance on all tasks.
\end{abstract}

\section{Introduction}
\label{sec:intro}
Powerful representations of data are useful in a number of ways. They improve model performance on small datasets by transferring data-driven insights from larger datasets. The models that create representations can also be used as pre-training for improved performance. If the model that generates the representation is non-reversible, then the representations can unlock applications in some privacy-sensitive scenarios. In this paper, we significantly improve state-of-the-art representations for paralinguistic speech tasks.

There are a number of promising data-driven speech representations. Some directions include self-supervised contrastive learning~\cite{trill, cola, jansen2017unsupervised}, predictive coding~\cite{unsupspeech, oord2019representation}, masked-unit prediction~\cite{hubert}, multi-task learning~\cite{pascual2019learning}, multimodal coincidence~\cite{arandjelovic2017look, jansen2019coincidence}, and intermediate representations from a supervised task~\cite{hershey2017cnn, asremb}. One of the most promising objectives for representation learning for speech recognition was proposed in the recent Wav2Vec 2.0~\cite{wav2vec2} framework, which combined Transformers~\cite{vaswani2017attention} and a self-supervised contrastive learning objective~\cite{oord2019representation}. The Wav2Vec 2.0 training objective was subsequently combined with more powerful Conformer architectures, producing large improvements in semi-supervised speech recognition applications~\cite{conformer, zhang2020pushing, bigssl}. This paper explores the use of these Conformer-based models to define fixed representations for non-ASR speech analysis and paralinguistics tasks.  To fully evaluate the potential of these models, we evaluate several model sizes and pretraining datasets combinations.

Recent work to establish a common benchmark has made it possible to directly compare speech representations~\cite{trill, superb}. In this work, we use the Non-Semantic Speech Benchmark (NOSS)~\cite{trill}, a collection of publicly available non-semantic speech tasks including speech emotion recognition, language identification, and speaker identification. Following \cite{frill}, we include masked speech detection~\cite{compare2020}. We also include three new tasks: synthetic speech detection~\cite{asvspoof}, an additional speech emotion recognition dataset~\cite{iemocap}, and dysarthria classification~\cite{euphonia}. Our work further establishes the usefulness of these embeddings over classical paralinguistic features, and can be used to improve other transfer-learning speech applications like voice imitation~\cite{jia2019transfer} and personalized ASR~\cite{dysarthric_asr}.

Finally, our work explores the impact of context window size on performance. We show that 2-second context windows are sufficient for nearly all tasks, but further context truncation can lead to large losses in performance.  Furthermore, we analyzed the range of embeddings produced by the sequence of Conformer blocks that define the encoder, demonstrating stable performance over a large portion of the network regardless of architecture complexity.  Using Centered Kernel Alignment (CKA) analysis~\cite{kornblith2019similarity,raghu2021vision}, we further demonstrate that the representations defined by this range of blocks are surprisingly similar, both within and (to lesser degree) across architectures.

The main contributions of this paper are:

 \begin{enumerate}[leftmargin=12pt]
 \setlength\itemsep{-1pt}
   \item Generate features for non-semantic speech tasks that set a new state-of-the-art (SoTA) performance on 7 of 9 tasks \textbf{using only time-averaged features and linear classification models}
   \item Analyze the performance versus context window size tradeoff, and show that 2-second context windows are sufficient
   \item Perform a more extensive embedding comparison than previously done, both in terms of downstream tasks and embeddings compared. Using a per-example analysis, we demonstrate that our embedding is strictly better than previous ones
   \item Demonstrate that similarly-performing representations in different architectures are similar in the CKA-sense
 \end{enumerate}

\section{Conformer-Based Representations}
\label{sec:representation}
\begin{table}[!t]
\vspace{-0.2cm}
\captionsetup{size=footnotesize}
\centering
\caption{Comparison of models. Resnetish50~\cite{hershey2017cnn}. MobileNetv3~\cite{mnetv3}. RNN-T~\cite{he2018streaming}. EfficientNet~\cite{tan2020efficientnet}. Conformer~\cite{conformer}. AudioSet~\cite{audioset}. YT-U~\cite{bigssl}, LL is Libri-Light~\cite{librilight}. $^*$``RA" stands for ``relative attention."}\label{tab:models}
\vspace{-3mm}
\footnotesize
\begin{tabular}{c c c c c} \toprule
  Name        & Architecture & Params & \begin{tabular}{@{}c@{}}Training \\ data \end{tabular} & \begin{tabular}{@{}c@{}}Labels \\ required \end{tabular} \\
  \midrule
  YAMNet~\cite{trill}        &   MobileNetv1    & 3.7M  & Audioset & Y \\
  TRILL~\cite{trill}         &   Resnetish50    & 24.5M & Audioset & N \\
  FRILL~\cite{frill}         &   MobileNetv3    & 10.1M & Audioset & N \\
  COLA~\cite{cola}           &   EfficientNetB0 & 4.0M  & Audioset & N \\
  ASR Emb~\cite{asremb}      &   RNN-T          & 122M  & -        & Y\\
  \midrule
  \begin{tabular}{@{}c@{}}Conformer XL \\ (No) RA$^*$ YT (LL) \end{tabular}
    &   Conformer      & 608M  
    & \begin{tabular}{@{}c@{}}YT-U \\ (LL) \end{tabular} & N \\
  \begin{tabular}{@{}c@{}}Conformer XXL \\ YT (LL) \end{tabular}
    &   Conformer      & 1.0B  
    & \begin{tabular}{@{}c@{}}YT-U \\ (LL) \end{tabular} & N \\
  Conformer G &   Conformer      & 8.0B  & YT-U & N \\
 \bottomrule
\end{tabular}
\vspace{-4.5mm}
\end{table}
\subsection{Architectures} 

Each of our proposed paralinguistic representations is defined using a \emph{speech encoder} comprised of a stack of convolution-augmented Transformer blocks known as Conformers~\cite{conformer}.  Each Conformer block inserts a small depthwise separable convolutional module between the Transformer's self-attention and MLP modules, which has been shown to be highly beneficial to many recognition applications.  The input to this speech encoder is the output of a 3-layer 1-dimensional convolutional \emph{feature encoder} that is applied to 80-bin log mel spectrogram features.  The spectrogams come from 16kHz audio that is resampled if necessary.  Two convolutional strides of two produce a vector time series that is downsampled by a factor of 4x, yielding a frame rate that is preserved throughout the entire speech encoder.

The models are trained using the Wav2Vec 2.0 contrastive loss~\cite{wav2vec2}: we
first extract encoded features from the feature encoder and then use masked features as inputs to the Conformer to create context vectors. These context vectors are trained to agree with the target context vectors, obtained by applying a linear layer to the initial encoded features, by a contrastive loss. Table~\ref{tab:models} lists the various Conformer architectures considered in our evaluation.  We consider three Conformer encoder complexities defined in the original study~\cite{bigssl}, including 608 million (24 layers/8 heads/1024D output), 1.0 billion (42 layers/8 heads/1024D output), and 8.0 billion parameters (36 layers/16 heads/3072D output). Also shown are corresponding details for five baseline representations that we include in our evaluation. These cover a range of model architectures, complexities, and training objectives.

\subsection{Pre-training Datasets}
We use two datasets for self-supervised training of the above architectures. The first is \textbf{YT-U}, a 900k hour dataset~\cite{bigssl} derived from YouTube. YT-U is built by first randomly collecting 3 million
hours of audio from ``speech-heavy" YouTube videos. The results are then segmented, and the non-speech segments are removed to yield approximately 900k hours of unlabeled audio data.

The second is \textbf{Libri-Light}~\cite{librilight}, which contains 60k hours of audio derived from open-source audio books in the LibriVox project. It is the largest publicly available, unlabeled semi-supervised dataset to date.

\section{Experiments}
\label{sec:experiments}
\subsection{Tasks: The Non-Semantic Speech Benchmark (NOSS)}

In order to fairly compare representations, we benchmark each representation on the same 9 tasks~(Table~\ref{tab:noss}). Our tasks include most of the original NOSS benchmark tasks~\cite{trill}, a mask-detection task used in representation benchmarking in \cite{frill}, a fake speech detection task~\cite{asvspoof}, an additional speech emotion recognition task~\cite{iemocap}, and a dysarthria classification task~\cite{euphonia}.  When a single scalar is necessary (e.g. to compare embeddings), we aggregate over the performances using the ``Aggregate Embedding Score", which is the average accuracy of a model, averaged across tasks.

\textbf{ASVSpoof2019}: We introduce the ASVSpoof2019~\cite{asvspoof} dataset as a new task in our benchmark. This task measures a model's ability to distinguish real from synthetic speech. We use the Logical Access (LA) portion of this dataset. The LA database contains bona fide and spoofed speech generated using 17 different text-to-speech and voice conversion systems. The task is especially challenging because spoofed speech in the test set is generated using techniques not seen in training.

\textbf{IEMOCAP}: The Interactive Emotional Dyadic Motion Capture database~\cite{iemocap} is an acted, multimodal, and multispeaker database. We use the improvised scenarios portion with categorical emotion labels. To compare fairly with previous SoTA work~\cite{superb}, we only use the audio component and only 4 of the 10 labels (angry, happy, neutral, and sad).

\textbf{Euphonia}: The Euphonia dataset~\cite{euphonia} is a large dysarthric speech dataset. Our task uses a 661 speaker subset of 29 identical phrases with manual dysarthria labels from speech-language
pathologists on their overall intelligibility using a five-point
Likert scale.

\begin{table}[!t]
\vspace{-0.2cm}
\captionsetup{size=footnotesize}
\centering
\caption{Downstream evaluation datasets.
$^*$Results in our study used a subset of Voxceleb filtered according to YouTube's privacy guidelines.}\label{tab:noss}
\vspace{-3mm}
\footnotesize
\begin{tabular}{c c c c c c } \toprule
  Dataset        & Target & Classes     & Samples     &  \begin{tabular}{@{}c@{}}Avg \\ length (s)\end{tabular}  \\
  \midrule
 VoxCeleb$^*$~\cite{voxceleb}  
 &  Speaker ID   & 1,251    & 12,052     & 8.4 \\
 VoxForge~\cite{voxforge}   
 & Language ID   & 6        & 176,438   & 5.8 \\
 \begin{tabular}{@{}c@{}}Speech \\ Commands\cite{speechcommands}\end{tabular} 
 & Command       & 12       & 100,503   & 1.0  \\
 \begin{tabular}{@{}c@{}}Masked \\ Speech~\cite{compare2020}\end{tabular}   
 & Mask wearing  & 2        & 36,554    & 1.0  \\
 ASVSpoof~\cite{asvspoof}   
 & \begin{tabular}{@{}c@{}}Synthetic \\ or not\end{tabular}
 & 2  & 121,461  & 3.2  \\
 Euphonia~\cite{euphonia}
 & Dysarthria & 5 & 15,224 & 6.4 \\
 CREMA-D~\cite{cremad}
 & Emotion       & 6        & 7,438     & 2.5 \\
 IEMOCAP~\cite{iemocap} 
 & Emotion       & 4        & 5,531         & 4.5    \\
 SAVEE~\cite{savee}
 & Emotion       & 7        & 480       & 3.8 \\
 \bottomrule
\end{tabular}
\vspace{-4.5mm}
\end{table}

\begin{centering}
\begin{table*}[t]
\vspace{-0.2cm}
\captionsetup{size=footnotesize}
\caption{Test performance on the NOSS Benchmark and extended tasks. ``Prev SoTA" are arbitrarily complicated models, but \textbf{all other rows are linear models on time-averaged input}.
$^\dagger$Filtered according to YouTube’s privacy guidelines. We omit previous SoTA results, since they used the entire dataset.
$^\ddagger$Task performance is reported using unweighted average recall~\cite{compare2020} instead of accuracy. Also, test set labels are not available, so we report accuracy on the eval set. 
$^{**}$Uses equal error rate~\cite{asvspoof}.
$^{\#}$The only non-public dataset. We exclude it from aggregate scores.
$^{\dagger\dagger}$Included in the table but not aggregate score, since it's less than 1/10th the size of the next smallest dataset and results have high variance.
$^*$Audio and visual features used in previous SoTA.
$^{+}$Prev SOTA performed cross-fold validation. We hold out speakers M05 and F05 as test.
$^{\small++}$YAMNet uses layer 10, as in~\cite{trill}.
$^\mathsection$Best per-task results are computed by taking the model/layer with the best results of the dev set, and reporting those results on the test set. If the dev set performance is better but the test results are worse, ``Best per-task" can be worse than ``Best overall".
}\label{tab:results}
\vspace{-0.3cm}
\footnotesize
\centering
\begin{tabular}{@{} l|ccccccccc @{}}
\toprule[2pt]
Model & 
Voxceleb1$^\dagger$ & 
Voxforge & 
\begin{tabular}{@{}c@{}}Speech \\ Commands\end{tabular}   &  
\begin{tabular}{@{}c@{}}Masked \\ Speech$^\ddagger$\end{tabular}  & 
\begin{tabular}{@{}c@{}}ASVSpoof\\ 2019$^{**}$\end{tabular} & 
Euphonia$^{\#}$ &
CREMA-D &
IEMOCAP &
SAVEE$^{\dagger\dagger}$
\\
\toprule[2pt]
\textbf{Prev SoTA}
& - 
& 95.4 \cite{sarthak2019spoken} 
& \textbf{97.9}~\cite{speech_commandssota} 
& 73.0~\cite{Szep2020} 
& 5.11~\cite{superb} 
& 45.9~\cite{asremb}
& 74.0$^*$ \cite{ghaleb2019} 
& 67.6$^{+}$~\cite{superb} 
& 84.0$^*$~\cite{savee} 
\\
\midrule
\textbf{Baselines} \\
\quad YAMNet$^{\small++}$ \cite{trill} 
& 10.9 & 79.8 & 78.5 & 59.7 & 9.23 & 43.0 & 66.4 & 57.5 & 69.2  \\
\quad TRILL \cite{trill}
& 12.6 & 84.5 & 77.6 & 65.2 & 7.46 & 48.1 & 65.7 & 54.3 & 65.0\\
\quad FRILL \cite{frill}   
& 13.8 & 78.8 & 74.4 & 67.2 & 7.45 & 46.6 & 71.3 & 57.6 & 63.3\\
\quad COLA \cite{cola}
& 11.7 & 71.0 & 60.6 & 65.0 & 4.58 & 47.6 & 69.3 & 63.9 & 59.2 \\
\quad ASR Emb \cite{asremb}
& 5.2 & 98.9 & 96.1  & 54.4 & 11.2 & \textbf{54.5} & 71.8 & 65.4 & 85.0 \\
\midrule
\textbf{Conformers} \\
\quad 
Best per-task$^\mathsection$ &
\textbf{53.5} & \textbf{99.8} & 97.5  & \textbf{74.2} & \textbf{2.5} & 53.6 & 87.2 & \textbf{79.2} & \textbf{92.5} \\
\quad  (model, layer \#) &
(XXL-YT, 25) & (G-YT, 19) & (CAP, 16) & (XL-LL RA, 5) & (CAP, 12) & (CAP, 13) & (G, 26) & (CAP, 15) & (CAP, 15) \\
\midrule
\quad 
\begin{tabular}{@{}c@{}}Best CAP per\\ task (layer \#)\end{tabular} &
50.3 (11) & 99.7 (14) & 97.5 (16) & 73.4 (10) & \textbf{2.5} (12)  & 53.6 (13) & \textbf{88.2}$^\mathsection$ (12) & \textbf{79.2} (15) & 92.5 (15) \\
\midrule
\quad 
\begin{tabular}{@{}c@{}}Best single\\ layer (CAP12)\end{tabular} &
51.0$^\mathsection$ & 99.7 & 97.0 & 68.9 & \textbf{2.5} & 51.5 & \textbf{88.2}$^\mathsection$ & 75.0 & 81.7\\
\bottomrule
\end{tabular}
\vspace{-2mm}
\end{table*}
\end{centering}

\subsection{Benchmark Results}

For our first set of experiments, we compute embeddings from our speech representation models~(Table \ref{tab:models}) and train simple models on the NOSS tasks using the same methodology from \cite{trill}. For each pair of benchmark task and embedding,  we train and evaluate a number of simple linear classification techniques (logistic regression,  balanced logistic regression, linear discriminant analysis) on top of clip-level average embeddings.  We choose the best performing classifier (as determined on dev set) and use it to report test performance for that (task, embedding) pair.  

\begin{figure}[t]
\vspace{-1mm}
\centering
  \includegraphics[width=0.9\columnwidth]{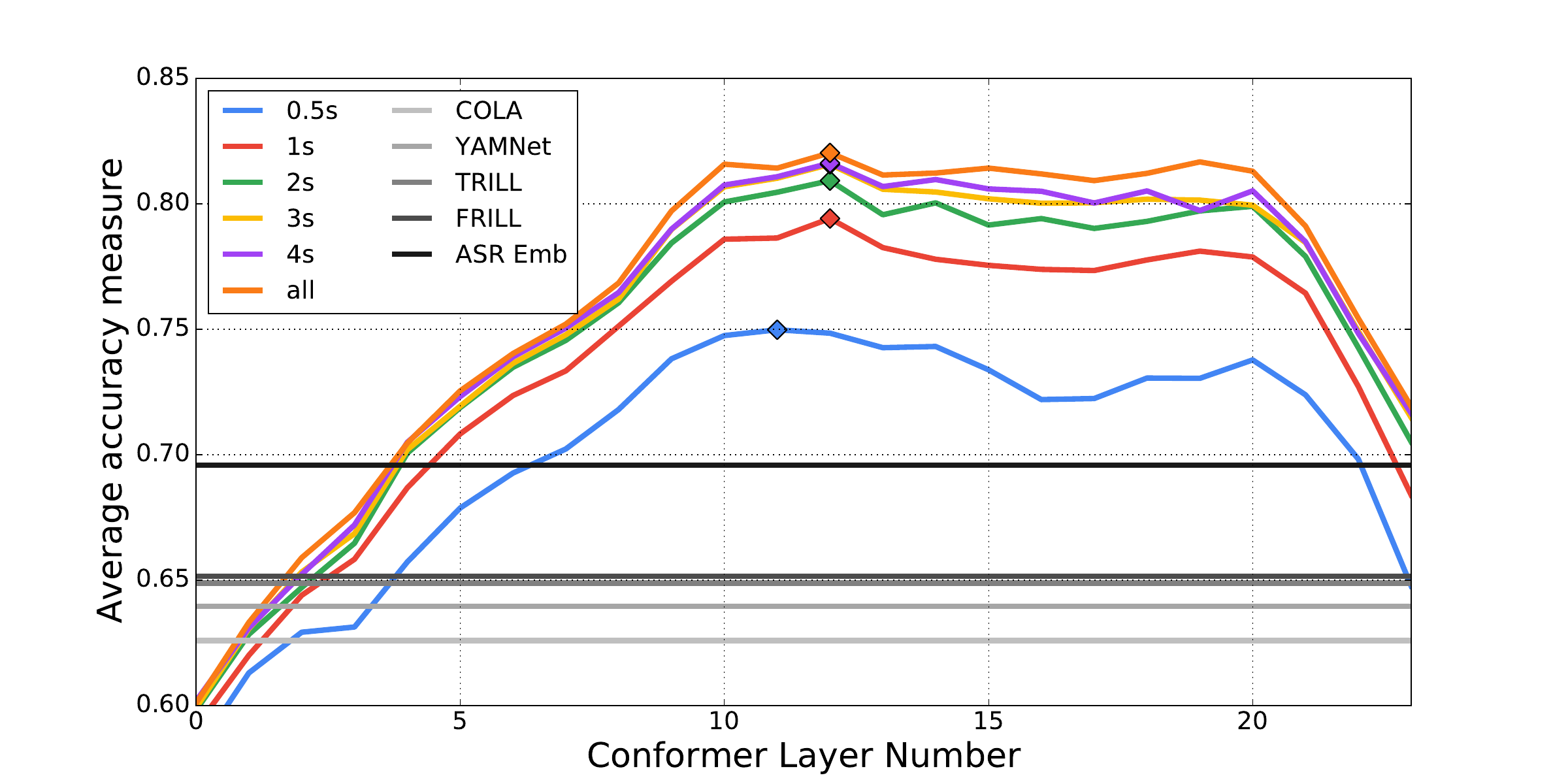}
  \includegraphics[width=0.9\columnwidth]{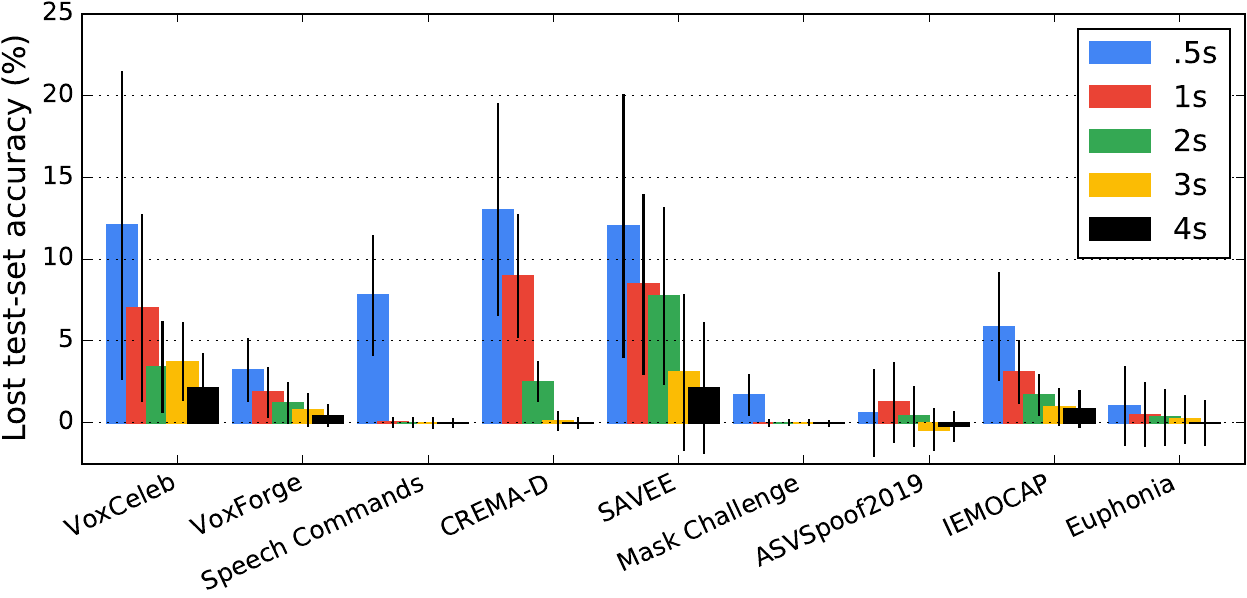}
  \vspace{-2mm}
  \captionsetup{size=footnotesize}
  \caption{\textbf{Upper)} Average test accuracy, averaged across tasks, for ``CAP." X-axis is the network layer. Different lines are different chunking values. \textbf{Lower)} Absolute accuracy lost due to smaller context windows. Error bars are 1 standard deviation. Each bar is a mean over (models) x (layers) = \textbf{192 values}.}\label{fig:context}
  \vspace{-3mm}
\end{figure}

Table~\ref{tab:results} shows the results of the benchmark.  Like recent studies we report the performance of the best (model, layer) pair on a per-task basis.  However, we also aim to establish a single universal set of features that serve all downstream tasks.  Thus, we also evaluate all intermediate representations and rank order them according to the Aggregate Embedding Quality on the dev set.  We then report performance on the test set in final line of Table~\ref{tab:results}.  It comes from layer 12/23 of the 600M parameter YT model, without relative attention. We call this model ``Conformer Applied to Paralinguistics," or ``\textbf{CAP}", and we refer to the best layer as ``\textbf{CAP12}." We note that this representation \textbf{was within 6\% accuracy of the per-task best layer on 7 of 9 tasks}.  Figure~\ref{fig:context} shows how the aggregate embedding quality varies in this model across intermediate layers.

\textbf{Linear classifiers on Conformer representations set a new SoTA on 7/9 tasks}: The ``best per-task" row in Table~\ref{tab:results} shows the test set results on the representations with the best dev-set performance. Linear models on these representations set a new SoTA on 7 of the tasks, often outperforming far more complex models. Furthermore, these linear models \textbf{outperform previous SoTA models that use more modalities than just speech} (CREMA, SAVEE).


\textbf{CAP12 significantly outperforms previous representations, especially on speech emotion recognition}: CAP12 outperforms every other non-Conformer representation on every dataset we used with the lone exception of "ASR Emb" on SAVEE.
Especially noteworthy are the results on CREMA-D and IEMOCAP, where \textbf{CAP12 outperforms previous embeddings by 16\% and 9\% respectively}.

\textbf{CAP12 significantly outperforms previous single-model SoTA on ASVSpoof2019}: Linear models on averaged CAP12 would've been the best single-model entry in the ASVSpoof2019 competition, and would've ranked 3rd overall~\cite{asvspoof}.


\begin{figure}[t]
{\centering
  \includegraphics[width=0.7\columnwidth]{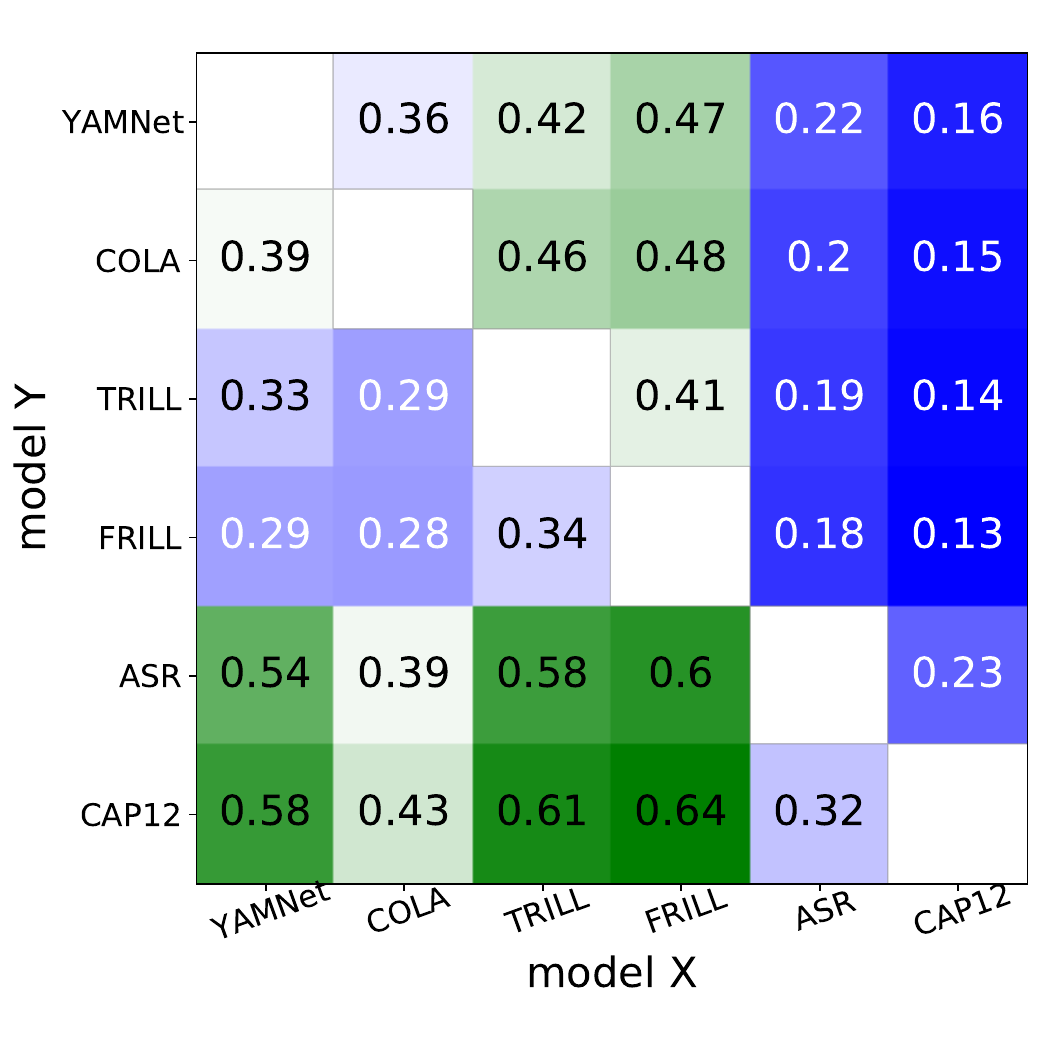}\hspace{0.2cm}
  \vspace{-4mm}
  \captionsetup{size=footnotesize}
  \caption{ 
  Each square is the probability that Model Y correctly predicts an example given that Model X and Model Y disagree on the prediction. The result is averaged over task. Each task is an average over examples.}\label{fig:comparison}}
  \vspace{-3mm}
\end{figure}

\textbf{CAP12 is strictly better than other representations}: Since aggregate performance ignores patterns of errors, we investigate the agreement between predictions made from different embeddings on a per-example basis.  Figure~\ref{fig:comparison} show that when CAP12 and other embeddings disagree, CAP12 is correct 32\%-64\% of the time, while other embeddings are correct only 13\%-23\% of the time. With the exception of the supervised ASR Embedding, it is relatively uncommon for other embeddings to be correct when CAP12 is wrong.

\subsection{Context window size}

Our second experiment studies the role of context window size.  Conformers, like Transformers, use the entire audio clip to generate embeddings, while CNN-based methods have fixed context-window sizes. To help understand how essential the large context window is for performance, we feed finite-window-sized inputs to the Conformer models, just like CNNs process input. We chunk the audio into fixed length sub-clips (e.g. 1 second), and have the Conformer model generate local embeddings independently.  We then average  over all local representations and evaluate the quality of the embedding on each NOSS task.  We also average the loss in performance across models, tasks, and layers to determine the average effect that finite-context windows have on downstream performance. Results are shown in Figure~\ref{fig:context}.

\textbf{2-second context windows are sufficient}: The best-performing layer for 4 / 3 / 2 / 1 / 0.5 second context windows are 99\% / 99\% / 98\% / 96\% / 91\% as accurate as the entire context window, averaged across tasks. This result is a function of the time-domain of the phenomena being studied, but is also a function of the fact that the datasets we use are known to include the signal we care about (average audio clips shown in Table~\ref{tab:noss}).




\subsection{Layerwise Analysis}

\begin{figure}[t]
\centering
  \includegraphics[width=0.92\columnwidth]{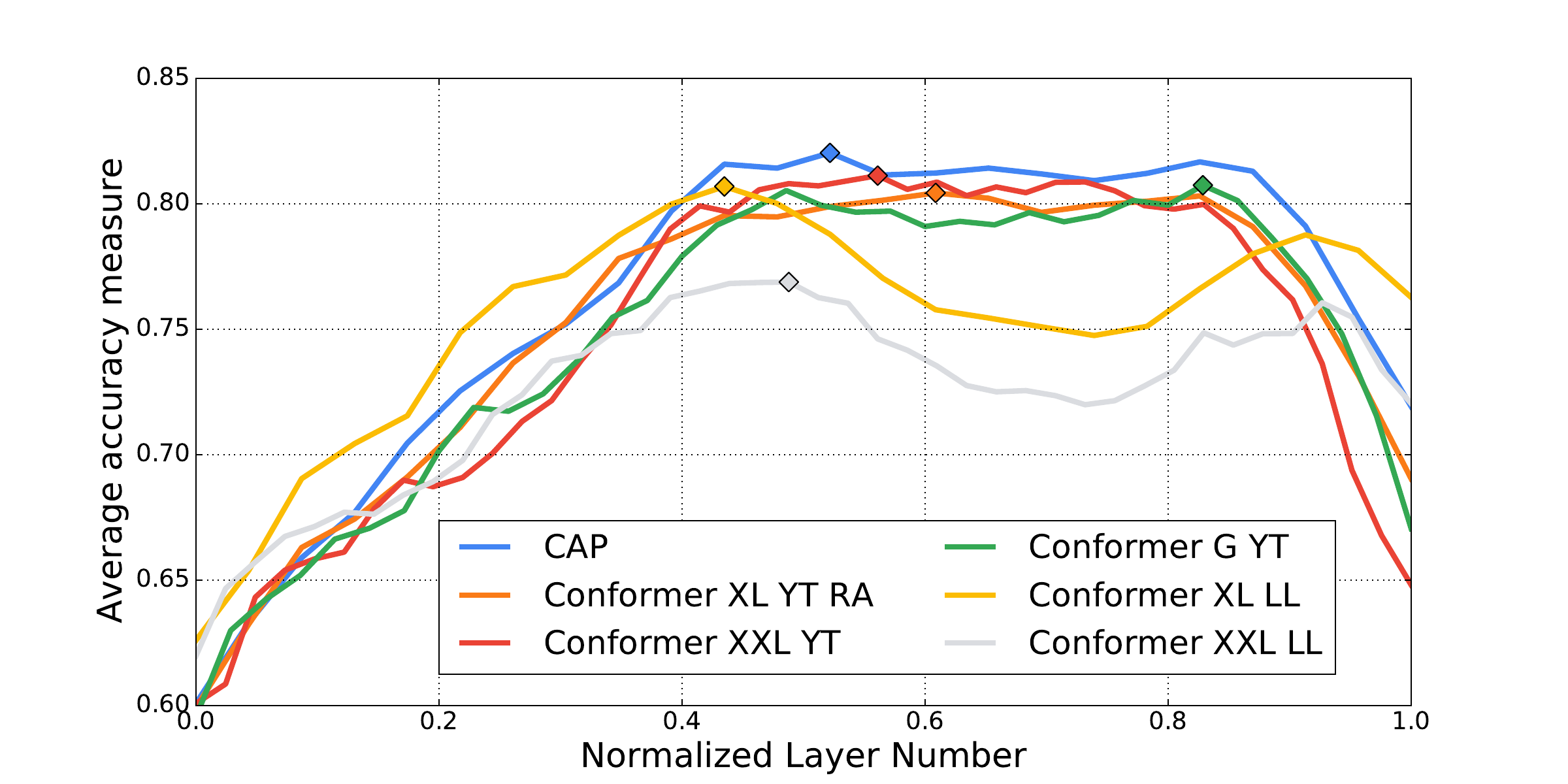}
  \includegraphics[height=0.43\columnwidth]{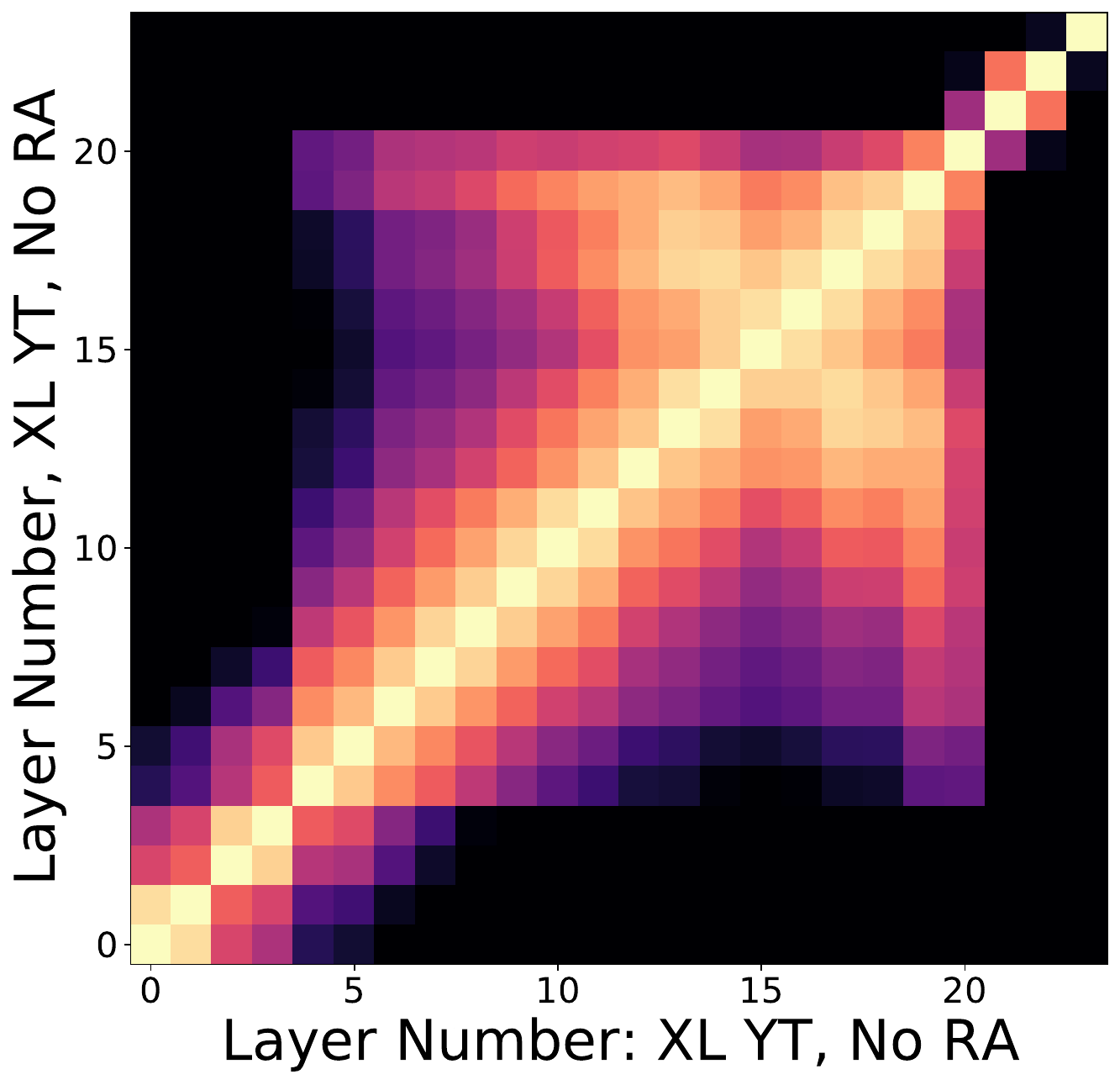} \hfill \includegraphics[height=0.435\columnwidth]{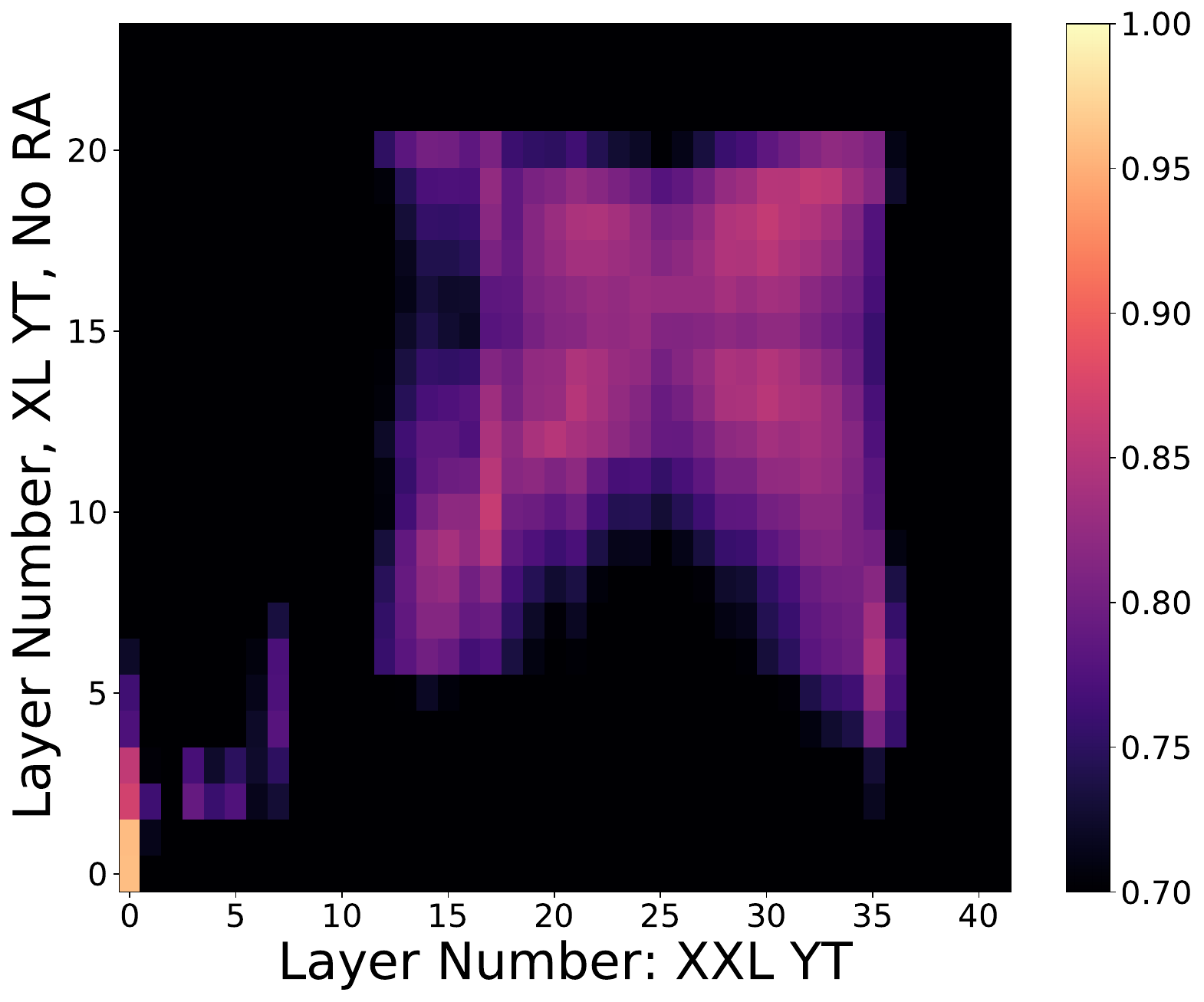}
  \includegraphics[height=0.47\columnwidth]{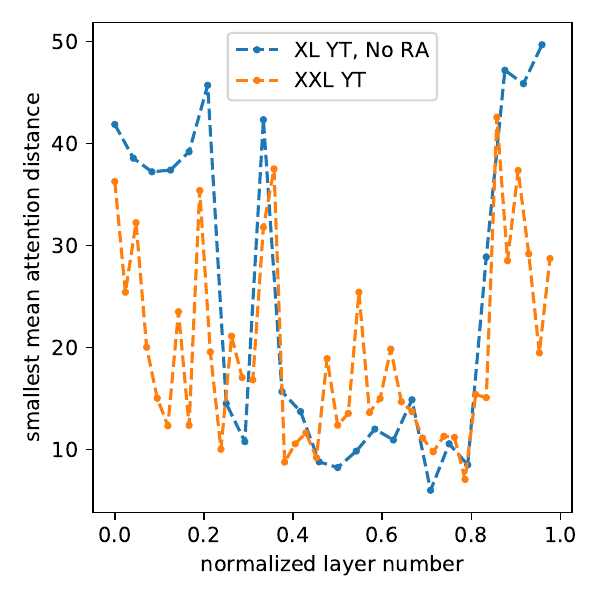} \hfill 
  \includegraphics[height=0.51\columnwidth]{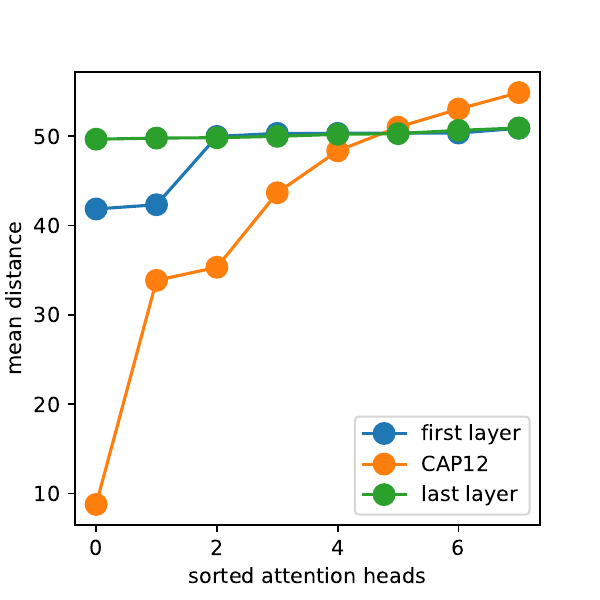}   
  \captionsetup{size=footnotesize}
  \caption{\textbf{Upper)} Average test-set accuracy, averaged across tasks, for 6 different Conformer models as a function of layer index normalized to [0.0, 1.0] using (layer \#) $/$ (\# of layers), where \# of layers is different for different models. \textbf{Middle)} Linear CKA scores between all pairs of layers: (left) within the Conformer XL YT network and (right) across the top performing Conformer XL YT and XXL YT networks. The colormap is truncated at 0.7 as is common to both images. \textbf{Lower)} Mean attention distance: (left) the shortest attention head on every layer; (right) all attention heads on 3 representative layers of the CAP model.}
  \label{fig:layerwise_analysis}
  \vspace{-3mm}
\end{figure}

The comparable performance of CAP12 relative to the optimal per-task embeddings identified in Table~\ref{tab:results} suggests a high degree of representational stability across layers and architectures evaluated.  Thus, our final set of experiments further probe the per-layer performance across layer and architecture.  Figure~\ref{fig:layerwise_analysis}(upper) plots the average accuracy on NOSS tasks as a function of layer for each architecture, but where layer index is normalized to a common [0, 1] scale.  We observe an overall dependence on pretraining dataset, with YouTube-trained models clearly outperforming LibriLight-trained ones.  However, within the YouTube models, we observe a surprisingly similar performance trajectory as we move through the normalized network position. Furthermore, for each of these models, we observe a wide performance plateau in the second half of each network. 

To test whether this behavior arises from representational similarity in the models' shared performance plateau, we apply linear Centered Kernel Alignment (CKA) between pairs of layers both within and across networks, following the methodology of a recent vision Transformer study~\cite{raghu2021vision}.  Briefly, CKA computes a $[0,1]$-valued similarity between two Gram matrices (using an arbitrary kernel function, which we take to be linear) separately computed from two representations over the same sample of input examples (see~\cite{kornblith2019similarity} for details).   Figure~\ref{fig:layerwise_analysis} (middle left) shows the pairwise layer similarity within the CAP network.  While each layer is most similar to its neighbors, we observe a large block of similar layers in the second half of the network corresponding to the performance plateau in Figure~\ref{fig:layerwise_analysis} (upper).  This indicates that the stable downstream performance is indeed fueled by a stable representation across these layers.  While the overall similarity across XL and XXL networks is lower in Figure~\ref{fig:layerwise_analysis} (middle right), we again see a block of similar layers corresponding to shared performance plateau.  This indicates similar characterization of paralinguistic properties in this stage of the network regardless of total network depth. 

Finally, in Figure~\ref{fig:layerwise_analysis} (lower right), we plot mean attention distances of self attention layers to study how much temporal context each layer is aggregating over. Following \cite{raghu2021vision}, we compute the mean attention distance as the attention probability-weighted average temporal distance for each attention head, and average over 1k clips from \cite{librilight}. We observe that higher and lower layers contain only global (long distance) attention heads, whereas middle layers have a mix of local and global ones. Interestingly, there is a clear correlation between the shortest attention distance on each layer (Figure~\ref{fig:layerwise_analysis}, lower left) and its average accuracy on NOSS tasks (Figure~\ref{fig:layerwise_analysis}, upper), which suggests the importance of local information for paralinguistic tasks.


\section{Conclusion}
\label{sec:conclusion}
In this paper, we introduce a class of Conformer-based self-supervised representation for speech. These representations set a new state-of-the-art performance on 7/9 paralinguistic speech tasks using only \textbf{embeddings averaged across time}, and using only \textbf{linear models on those embeddings.} Furthermore, these representations substantially outperform other speech representations despite not using labels for training. Even though the models use the entire context window to generate embeddings, we demonstrate that 2-second windows give 96\% the performance of the full context window on 7 of 9 tasks, and that \textbf{these representations with 500ms context windows still outperform previous representations}. Finally, we show that Conformer models of different sizes and datasets learn comparable representations at similar parts of the network, indicating that our findings are fundamental to the problem and not a superficial artifact.

\bibliographystyle{IEEE}
\bibliography{biblio}

\end{document}